\documentclass{article}
\usepackage{arxiv}
\usepackage[utf8]{inputenc} 
\usepackage[T1]{fontenc}    
\usepackage{hyperref}       
\usepackage{url}            
\usepackage{booktabs}       
\usepackage{amsfonts}       
\usepackage{nicefrac}       
\usepackage{microtype}      
\usepackage{lipsum}
\usepackage{amsmath,amssymb,graphicx}

\title{Mathematical methods of diagonalization of quadratic forms applied to the study of stability of thermodynamic systems}

\author{
  F. N. Lima\thanks{I am corresponding author.} \\
  GTMCOQ, Instituto Federal do Piau\'i -- IFPI \\
  São Raimundo Nonato, Piau\'i, 64770-000, Brazil.\\
   \texttt{nogueira@ifpi.edu.br} \\
   \And
 J. M. De Sousa \\
  GTMCOQ, Instituto Federal do Piau\'i -- IFPI \\
  São Raimundo Nonato, Piau\'i, 64770-000, Brazil.\\
   \texttt{josemoreiradesousa@ifpi.edu.br} \\
  }

\begin{document}
\maketitle

\begin{abstract}
In this paper, we use quadratic forms diagonalization methods applied to the function thermodynamic energy to analyze the stability of physical systems. Taylor's expansion was useful to write a quadratic expression for the energy function. We consider the same methodology to expanding the thermodynamic entropy and investigate the signs of the second-order derivatives of the entropy as well as previously to the thermodynamic energy function. The signs of the second-order derivatives to the Helmholtz, enthalpy and Gibbs functions are also analysed. We show the immediate consequences on the stability of physical systems due to the signs or curvatures of the second-order derivatives of these thermodynamic functions. The thermodynamic potentials are presented and constructed pedagogically as well as demonstrated the main mathematical aspects these surfaces. We demonstrate the power of superposition of mathematical and physical aspects to understand  the stability of thermodynamic systems. Besides, we provide a consistent mathematical demonstration of the minimum, maximum, and saddle conditions of the potentials. We present here a detailed approach on aspects related to the curvature of the thermodynamic functions of physical interest with consequences on stability. This work can be useful as a part or supplement material of the traditional physics curriculum that requires a solid formation in thermodynamics, particularly about formal aspects on the stability.
\end{abstract}


\section{Introduction}

In real physical systems the great number of atoms, molecules or ions, of the order of the Avogadro's number becames the measuring of the system's energy a non trivial problem. The microscopically comprehension of these systems require considering the almost infinite fredom degrees envolved as well as the atomic coordinates of the components of the system on time, etc. Fortunately, a great number of these freedom degrees are eliminated by considering statistical averages and not manifested in a macroscopic level. However, these modes can be reponsible for energy transfer in heat form. It would be impossible in practical situations computing the total energy of a system considering all freedom degrees dependent on time. But there is a theory independent on time, namely, thermodynamics that can be used to determine the macroscopic equilibrium of physical systems. Accordingly physics, can be extracted by constructing an appropriate thermodynamic physical function \cite{callen1998thermodynamics,de2005termodinamica,wreszinski2003termodinamica}.

Currently, we know that physical laws are unaltered in relation to time. This is intimately related to one of the most important principles in Physics, that is, the principle of energy conservation \cite{feynman2011feynman, callen1998thermodynamics}. The energy conservation  has been exaustively tried and confirmed along of the ultimate years. We should understand a physical system as an agglomerate of complex components, atoms, molecules, ions, etc., interacting with complex forces rigorously defined by states where the principle of energy conservation is valid \cite{harman1982energy,perkins2000introduction,shrimpton2009charge, callen1998thermodynamics}.Thus, it is intuitive considering a well defined energy function to describe the physical macroscopic properties of thermodynamical systems. Besides, in reason of the complexity in measuring the total energy of the systems it is convenient to assume some state whereby the energy is arbitrary defined as zero and measuring the total energy in connection that state \cite{callen1998thermodynamics,de2005termodinamica,wreszinski2003termodinamica}. 

In practice, it is possible measure only the differences of energy. The energy of interest in physics that can be used to describe the thermodynamic physical properties is defined here as $u(s,v)$, and it is a function on $s$ entropy and $v$ volume. In this case, we are considering a system with only one component the fundamental equation can be written in the form $u=u (s, v)$, where the number of moles $N$ is iserted in the variables of energy, entropy, and volume. It is important to remember that the postulate of minimum energy has empirical nature (see Ref. \cite{callen1998thermodynamics}). This function is defined to all equilibrium states whereby are valid the princicle of energy conservation and under the condition of the solid princicle of minimum energy. Whenever we invoke the thermodynamical energy function of the system, we are referring to the internal energy $u=u(s,v)$ defined in relation to a state taken arbitrarily as zero.

There is an equivalent approach in thermodynamics that can be used the same way to describe the physical properties of systems, in this case, through another relevant physical function known as entropy $s=s(u,v)$. The entropy $s=s(u,v)$ is a function on $u$ energy and $v$ volume. The entropy function $s(u,v)$ is valid under the princicle of maximum entropy, being also of empirical nature, as the princicle of minimum energy. The same way of $u(s,v)$, for the entropy function $s=s(u,v)$ we are considering a system of one component \cite{callen1998thermodynamics,de2005termodinamica,wreszinski2003termodinamica}.

In the last decades, the thermodynamic theory allowed the investigation of physical quantities of interest for thermodynamic systems in several areas of applied sciences. It has been possible to study thermodynamic quantities and molecular mobility values for several amorphous compounds and in investigating its crystallization behavior \cite{zhou2002physical}. The study of phase equilibrium thermodynamics is useful in the progress of chemical engineer \cite{zeck1991thermodynamics}. There are many pratical and applied problems where the  thermodynamic theory is essential, such as to better undertand of dynamics in thermodynamics of binding, in advanced study of batteries, in estimation of solvation parameters for monatomic ions, new values for the Gibbs energy, enthalpy, and entropy of solvation of monatomic ions in water, in computing of relative thermodynamic stabilities of amino acids, in advanced methods of molecular dynamics, and others applications \cite{rapaport2004art,allen2017computer,o1990thermodynamic,fawcett1999thermodynamic,forman1999dynamics}.

In all theoretical and experimental applications of thermodynamic theory an appropriate knowledge of the postulational  thermodynamic is essential. The postulational approach is relevant for the correct understanding of a fundamental point of view on physical theories that leads to the thermal, and mechanical stability of physical systems. In this paper, we obtain the thermal, and mechanical stability directly of the curvature or signs of the thermodynamic potentials, including the energy function. In addition, we demonstrate by application of a methodology of quadratic forms diagonalization to reveal the signs of the second-order derivatives of the several thermodynamic functions. What is more, the conclusions about the signs of the physical quantities of interest related to the stability emerge as immediate consequences of the signs (or curvature) of the mentioned thermodynamic functions. Taylor's series appear as a pedagogical tool to obtaining the quadratic forms for the used functions.

This work is organized as follows: In section \ref{diagonalization}, we present the diagonalization of the energy function as well as the signs of second-order derivatives of the energy. We also present the same way the signs of other functions entropy, Helmholtz, enthalpy, and Gibbs. In section \ref{review}, we provide a review on interest physical quantities. In section \ref{consequences}, we show the immediate consequences on the stability of physical systems originated by curvature of physical systems previously presented as well as the signs of the second-order derivatives. As it turns, we summarize the main results of the power of superposition of mathematical and physical aspects to understand the stability of thermodynamic systems as well as on aspects related to the curvature of thermodynamic functions of physical interest with consequences on the stability.

\section{Diagonalization of the thermodynamic energy function\label{diagonalization}}

\begin{figure}[!h]
	\begin{center}
		\centerline{\includegraphics*[width=0.47\textwidth]{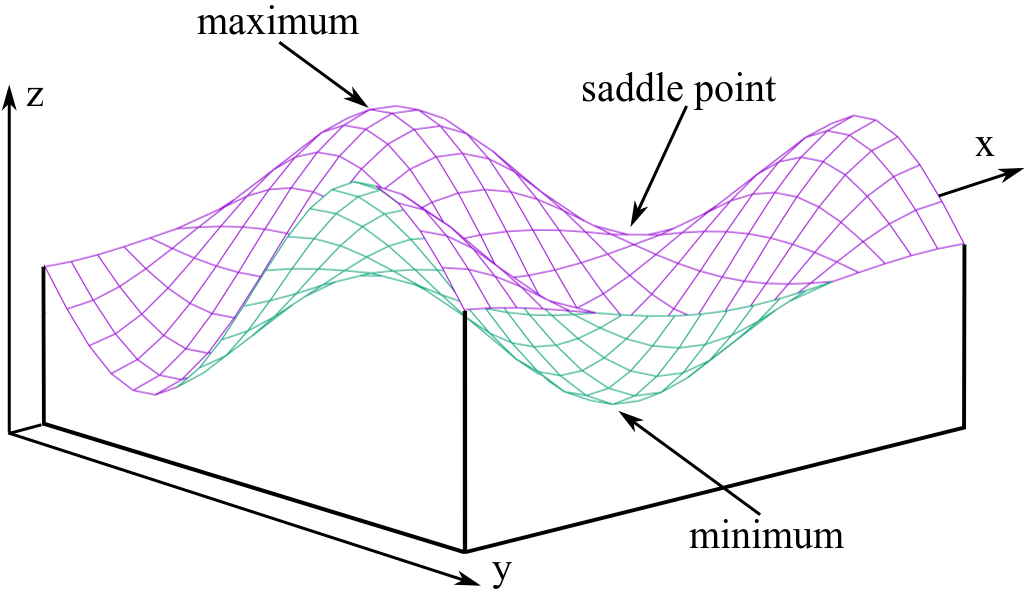}}
		\caption{Schematic view for minimum, maximum and saddle arbitrary surfaces.} \label{fig1}
	\end{center}
\end{figure}

We shall assume the existence of a well-defined thermodynamic energy function $u(s,v)$ capable of describing the thermodynamic macroscopic properties of physical systems, as discussed in the introduction. In particular, we are interested here in computing the signs (related with the stability) of the physical quantities reviewed in section \ref{review}, i. e., specific heats and compressibility coefficients. The energy function is defined as a function of entropy and volume. Let us begin expanding the energy function $u(s,v)$ in Taylor's series until second-order approximation as described below \cite{riley2006mathematical,apostol2007calculus,arfken1999mathematical}. 
\begin{flalign} 
u(s,v)=u(s_{0},v_{0})+\frac{\partial u}{\partial s}(s_{0},v_{0})(s-s_{0}) \nonumber \\
+ \frac{\partial u}{\partial v}(s_{0},v_{0})(v-v_{0}) \nonumber \\
+\frac{1}{2!}\Bigg[\frac{\partial^{2} u}{\partial s^{2}}(s_{0},v_{0})(s-s_{0})^{2}\nonumber \\
+\frac{\partial^{2} u}{\partial v^{2}}(s_{0},v_{0})(v-v_{0})^{2}\nonumber \\
+ 2\frac{\partial^{2} u}{\partial s\partial v}(s_{0},v_{0})(s-s_{0})(v-v_{0})\Bigg] + ... 
\label{eq:1}
\end{flalign}
Using the physical principle of minimum energy at stationary point $(s_{0},v_{0})$ ($d^2u\geq0$, see \cite{callen1998thermodynamics}) and by analogy with the one-variable differential calculus, it is possible to simplify the  expression above making:
\begin{flalign} 
\frac{\partial u}{\partial s}(s_{0},v_{0}) = 0 \textrm{ and } \frac{\partial u}{\partial v}(s_{0},v_{0}) =0.
\label{eq:2}
\end{flalign}
These conditions also occurs at stationary points of maximum and saddle points. Fig. \ref{fig1} shows a schematic view for minimum, maximum and saddle arbitrary surfaces \cite{riley2006mathematical,apostol2007calculus,arfken1999mathematical}. 

Here, the main goal is not to produce a broad discussion about the thermodynamic  energy, entropy function or other physical functions that are presented follow, in terms of its extensive and intensive physical parameters \cite{callen1998thermodynamics}. However, we want to discuss the mathematical features of minimum, maximum and saddle surfaces of the thermodynamic functions in order to focus on stability of physical systems. Although the mathematical aspects of these functions have been obtained from thermodynamic principles  of minimum energy or equivalently maximum entropy. Yet, the mathematics presented here is valid to the independent and arbitrary functions, as long as they meet certain conditions of derivation and continuity. A complete review on thermodynamics functions discussed in this section and its extensive or intensive parameters is found in Ref \cite{callen1998thermodynamics,de2005termodinamica,wreszinski2003termodinamica}.

Resuming the previous discussion, and after the simple reordered of terms in Eq. \ref{eq:1}, we can obtain 
\begin{flalign} 
2![u(s,v)-u(s_{0},v_{0})]\cong \frac{\partial^{2} u}{\partial s^{2}}(s_{0},v_{0})(s-s_{0})^{2} \nonumber \\
+\frac{\partial^{2} u}{\partial v^{2}}(s_{0},v_{0})(v-v_{0})^{2} \nonumber \\
+ 2\frac{\partial^{2} u}{\partial s\partial v}(s_{0},v_{0})(s-s_{0})(v-v_{0}) 
\label{eq:3}
\end{flalign}

Let us define $\tilde{u}(s,v) \equiv 2![u(s,v)-u(s_{0},v_{0})]\equiv 2!\Delta u$, $a \equiv \frac{\partial^{2} u}{\partial s^{2}}(s_{0},v_{0})$, $b \equiv \frac{\partial^{2} u}{\partial v^{2}}(s_{0},v_{0})$, $c \equiv 2\frac{\partial^{2} u}{\partial s\partial v}(s_{0},v_{0})$, $\Delta s=s-s_{0}$, $\Delta v=v-v_{0}$ and rewriting the Eq. \ref{eq:3} as follows

\begin{flalign} 
\tilde{u}(s,v) \cong a(\Delta s)^{2} +b(\Delta v)^{2}+c(\Delta s)(\Delta v). 
\label{eq:4}
\end{flalign}

Observe that the right side of Eq. \ref{eq:4} is a quadratic form and the left side $\tilde{u}(s,v)$ can be interpreted the same way as the thermodynamic energy function but mathematically multiplied and suppressed by the constants $2!$ and $u(s_{0},v_{0})$, respectively. Then, there is no physical difference between studying the sign of $u(s,v)$, and $\tilde{u}(s,v)$. The principle of minimum energy is also valid to the function $\tilde{u}(s,v)$, because this function is the original energy function, at less than a multiplicative constant, and additive.

It is important to remember the matricial form of the quadratic expression given by Eq. \ref{eq:4} (see Ref \cite{riley2006mathematical, anton2001algebra} for a general discussion on quadratic forms)
\begin{eqnarray}
\tilde{u}(s,v)=
\begin{pmatrix}
\Delta s\ \Delta v
\end{pmatrix}
\begin{pmatrix}
a & c\\
c & b
\end{pmatrix} 
\begin{pmatrix}
\Delta s\\
\Delta v
\end{pmatrix}=\Delta x^{T}M\Delta x,
\label{eq:5}
\end{eqnarray}
where $\Delta x$ is a column vector with $\Delta s$ and $\Delta v$ components, and $\Delta x^{T}$ its transpose of $\Delta x$. See that $M$ is a symmetric matrix and it is convenient to use a matrix diagonalization procedure to write $\tilde{u}(s,v)$ in a canonical form $\tilde{u}(s,v)=\Delta x^{'T}D\Delta x{'}$ to study the sign of $\tilde{u}(s,v)$ more easily. Here, $D$ is the eigenvalues matrix of $M$ with $\lambda_{1}$, $\lambda_{2}$ components, and the $\Delta x^{'}$ is the column eigenvector (with  $\Delta s^{'}$ and $\Delta v^{'}$ components) of the diagonal matrix $D$ as well as $\Delta x^{'T}$ its transpose.  

The canonical form to $\tilde{u}(s,v)$ can be expressed by Eq. \ref{eq:6}. A complete review on quadratic forms diagonalization can be found in Ref. \cite{anton2001algebra}.
\begin{eqnarray}
\tilde{u}(s,v)= \Delta x^{'T}D\Delta x^{'}= \nonumber\\
\begin{pmatrix}
\Delta s^{'}\ \Delta v^{'}
\end{pmatrix}
\begin{pmatrix}
\lambda_{1} & 0\\
0 & \lambda_{2}
\end{pmatrix} 
\begin{pmatrix}
\Delta s^{'}\\
\Delta v^{'}
\end{pmatrix}=\lambda_{1}\Delta s^{'2}+\lambda_{2}\Delta v^{'2}
\label{eq:6}
\end{eqnarray}
The physical principle of minimum energy imposes the function $\tilde{u}(s,v)$ in Eq. \ref{eq:6} the following mathematical condition $\tilde{u}(s,v)=\lambda_{1}\Delta s^{'2} + \lambda_{2}\Delta v^{'2}\geq0$ that occurs only when $\lambda_{1}\geq0$ and $\lambda_{2}\geq0$ (for any sets of values of $\Delta s^{'}$ and $\Delta v^{'}$, see \cite{riley2006mathematical, anton2001algebra}). In order to obtain the eigenvalues $\lambda_{1}$ e $\lambda_{2}$, we need to diagonalize $M$ by solving the equation below
\begin{eqnarray}
\begin{vmatrix}
a-\lambda & c\\
c & b - \lambda
\end{vmatrix}=0. 
\label{eq:7}
\end{eqnarray}

\begin{flalign} 
(a-\lambda)(b-\lambda) - c^{2}=0
\label{eq:8}
\end{flalign}

\begin{flalign} 
ab-(a+b)\lambda +\lambda^{2} - c^{2}
\label{eq:9}
\end{flalign}

\begin{flalign} 
\lambda^{2} -(a+b)\lambda + ab - c^{2}
\label{eq:10}
\end{flalign}

\begin{flalign} 
\lambda_{\pm}=\frac{(a+b)\pm \sqrt{(a+b)^{2}-4(ab - c^{2}})}{2}
\label{eq:11}
\end{flalign}

Remembering that $a$, $b$, and $c$ coefficients were defined as 
\begin{flalign} 
a \equiv \frac{\partial^{2} u}{\partial s^{2}}\ \textrm{, } b \equiv \frac{\partial^{2} u}{\partial v^{2}}\ \textrm{ and } c \equiv 2\frac{\partial^{2} u}{\partial s\partial v}.
\label{eq:12}
\end{flalign}

As discussed above the minimum energy principle imposes the thermodynamic energy function $\tilde{u}(s,v)\geq0$ (for all sets of values of $\Delta s^{'}$ and $\Delta v^{'}$) that require in Eq. \ref{eq:6} for the eigenvalues of $M$, $\lambda_{1}\geq0$ and $\lambda_{2}\geq0$. Observe that, in Eq. \ref{eq:11} for both eigenvalues $\lambda_{\pm}$ to be positive (corresponding to a minimum), we need $a$ and $b$ positive, and besides $(a+b)\pm \sqrt{(a+b)^{2}-4(ab - c^{2}})$ also positive. Here, by simplification of this end requisite, we have $ab - c^{2}>0$. It is important to emphasize that $a=\frac{\partial^{2} u}{\partial s^{2}}$ and $b=\frac{\partial^{2} u}{\partial v^{2}}$ positive were already expected, by analogy with the calculus of the one-variable functions, where the second-order derivatives are positive in the minimum condition. All these mathematical requisites provide 
\begin{flalign} 
\frac{\partial^{2} u}{\partial s^{2}}\geq0\ \textrm{, } \frac{\partial^{2} u}{\partial v^{2}}\geq0 \textrm{ and }
\frac{\partial^{2} u}{\partial s^{2}}\frac{\partial^{2} u}{\partial v^{2}} - \Bigg(\frac{\partial^{2} u}{\partial s\partial v}\Bigg)^{2}\geq0.
\label{eq:13}
\end{flalign}

We have so far directly applied and also proven an important mathematical theorem (see Refs. \cite{riley2006mathematical,apostol2007calculus,arfken1999mathematical}). Accordingly, Taylor's series has been utilized in the function thermodynamic energy to show relevant mathematical aspects of the second-order derivatives of the thermodynamic energy function (under the imposed physical condition of minimum energy), as seen in Eq. \ref{eq:13}. In the next section, we briefly address a review on some quantities of physical interest, as specific heat, coefficient of thermal expansion, etc. These physical quantities are defined in terms of derivatives of $\tilde{u}(u,v)$, like those that appear in Eq. \ref{eq:13}. As it turns, we can better understand the signs of the physical quantities as well as the consequences on thermodynamic stability that are presented in section \ref{consequences}. But before that, it is plausible to discuss another important physical function, i. e., the entropy. 

Using the same procedure above applied to the energy function $\tilde{u}(s,v)$, it is possible to define a function thermodynamic entropy $\tilde{s}(u,v)$. Physically this function has a maximum, as we know from the principle of maximum entropy as well as it is a function on energy and the volume. Then, the following second-order derivatives for $\tilde{s}$ are valid, and can be obtained the same way as for $\tilde{u}$
\begin{flalign} 
\frac{\partial^{2} s}{\partial u^{2}}\leq0\ \textrm{, } \frac{\partial^{2} s}{\partial v^{2}}\leq0 \textrm{ and }
\frac{\partial^{2} s}{\partial u^{2}}\frac{\partial^{2} s}{\partial v^{2}} - \Bigg(\frac{\partial^{2} s}{\partial u\partial v}\Bigg)^{2}\geq0.
\label{eq:14}
\end{flalign}

The first two expressions in Eq. \ref{eq:14} are already expected of one-variable calculus, but the third one is not. This last inequality is obtained by the diagonalization of the quadratic given by $\tilde{s}(u,v)$, if we take the same procedure of Taylor's expansion used to obtain $\tilde{u}(s,v)$ in Eq. \ref{eq:4}, where the canonical form to $\tilde{s}(u,v)$ can be expressed by Eq. \ref{eq:6s}, and it is obtained the same way as $\tilde{u}(s,v)$ (see Eq. \ref{eq:6})
\begin{eqnarray}
\tilde{s}(u,v)= \Delta x^{'T}D\Delta x^{'}=\lambda_{1}\Delta u^{'2}+\lambda_{2}\Delta v^{'2},
\label{eq:6s}
\end{eqnarray}
with $D$ being a diagonal matrix of eigenvalues that appears in the qua\-dratic form of $\tilde{s}(u,v)$, as discussed on $\tilde{u}(s,v)$. The imposition of negative $\lambda_{1}$ and $\lambda_{2}$ in Eq. \ref{eq:6s} due to the principle of maximum entropy leads to the set of inequalities given by Eq. \ref{eq:14}, without any necessity of analogy with one-variable calculus. Note that the imposition of signals in the eigenvalues leads naturally to Eq. \ref{eq:14}. Let us refer to the sign of the second-order derivatives of energy, entropy or to the other thermodynamic function of generic way as curvature (or concavite) of the function treated in relation to some variable.  Although $u$ is known as a convexity function while $s$ is known as a concave function (see more details in Ref. \cite{callen1998thermodynamics}).

Later we see that second-order derivatives in Eq. \ref{eq:13} or equivalently in Eq. \ref{eq:14} are associated to the physical quantities of interest. Due to the minimum energy principle as well as the maximum entropy principle, the second-order derivatives in energy and in entropy have in its signs the caracteristics discussed above. Therefore, the mentioned physical quantities have well defined signs intimatily related to the thermodynamic stability of physical systems. We have the same physical consequences on stability independent on the principle used (see more details in Ref. \cite{callen1998thermodynamics,de2005termodinamica,wreszinski2003termodinamica}).

In more realistic physical problems, it is convenient to work with the thermodynamic potentials of Helmoltz, enthalpy and Gibbs instead of thermodynamic energy. Thus, using Legendre transformations, we can change the extensive variables, or part of them, in the function thermodynamic energy by the intensive variables, making the appropriate Legendre transformation. For a broad discussion on extensive and intensive thermodynamic variables see Ref. \cite{callen1998thermodynamics}. Legendre's transformation is a process of change of variables, described below directly for the energy function. It is possible to write $u(s,v)$ as
\begin{flalign} 
du(s,v)=\frac{\partial u}{\partial s}ds+\frac{\partial u}{\partial v}dv,
\label{eq:15}
\end{flalign}
where the temperature is defined by $T\equiv \frac{\partial u}{\partial s}$ with $v$  constant, and the pressure is defined by $p\equiv - \frac{\partial u}{\partial v}$ with $s$ constant. Rewriting Eq. \ref{eq:15} with these definitions, we have
\begin{flalign} 
du=Tds-pdv.
\label{eq:16}
\end{flalign}

In order to introduce another energy function that instead of being a function of $s$ and $v$ it is expressed in terms of $T$ and $v$, it is necessary to take a Legendre transformation (change $s$ by $T$) in extensive parameter $s$. To do that, we introduce the intensive parameter $T$ changing the mentioned variables as described below
\begin{flalign} 
d(Ts)=Tds-sdT\nonumber \\
Tds=d(Ts)-sdT.
\label{eq:17}
\end{flalign}

Substituting Eq. \ref{eq:17} into Eq. \ref{eq:16} and some simple manipulations
\begin{flalign} 
du=d(Ts)-sdT-pdv\nonumber \\
d(u-Ts)=-sdT-pdv\nonumber\\
df=-sdT-pdv,
\label{eq:18}
\end{flalign}
where 
\begin{flalign} 
f\equiv u -Ts.
\label{eq:19}
\end{flalign}

The energy function defined by Eq. \ref{eq:19} it is known as Helmholtz free energy $f$. A general treatment on Legendre transformation can be found in Ref. \cite{boas2006mathematical}. This free energy is a function on $T$ and $v$, $f=f(T,v)$, as can seen in Eq. \ref{eq:18}. It is important to clarify that free energy $f$ defined as function of $T$ and $v$ has modified its concavite in relation to the new introduced parameter by Legendre transformation in $s$, i. e., the second-order derivatives of $f$ on $T$ is negative now ($\frac{\partial^{2} f}{\partial T^{2}}\leq0$). It is not hard to show that Legendre transformation change the curvature (in relation to the introduced parameter T) of the new "transformed energy" and defined by $f$ potential. The energy function has the curvature in relation to the $s$ variable demonstrated in Eq. \ref{eq:13}, $\frac{\partial^2 u}{\partial s^2}\geq0$, because $u$ is a surface of minimum. It is sufficient to elucidate that $\frac{\partial^2 f}{\partial s^T}\leq0$, and so $f$ has a maximum in relation to the introduced variable $T$. On other words, we need to demonstrate that $\frac{\partial^2 u}{\partial s^2}$ and $\frac{\partial^2 f}{\partial T^2}$ have opposite signs.  As it is better discussed in section \ref{consequences} the Eq. \ref{eq:18} provides $-s=\frac{\partial f}{\partial T}$ at $v$ constant. Taking the derivation of this equation side by side in relation to $T$ keeping $v$ constant, we have $-\frac{\partial s}{\partial T}=\frac{\partial^2 f}{\partial T^2}$, where the right side represents the curvature of the Helmholtz potential relative to $T$. Remembering that temperature definition is given by $T=\frac{\partial u}{\partial s}$ at $v$ constant (see Eqs. \ref{eq:15} and \ref{eq:16}) as well as by derivation of both members in the definition of temperature in relation to $s$ at $v$ constant, $\frac{\partial T}{\partial s}=\frac{\partial^2 u}{\partial s^2}$. Then, by comparing this expression with $-\frac{\partial s}{\partial T}=\frac{\partial^2 f}{\partial T^2}$ (or equivalently $-1/\frac{\partial T}{\partial s}=\frac{\partial^2 f}{\partial T^2}$) it is possible to conclude $-1/\frac{\partial^2 u}{\partial s^2}=\frac{\partial^2 f}{\partial T^2}$ (or equivalently $\frac{\partial^2 u}{\partial s^2}\frac{\partial^2 f}{\partial T^2}=-1$). This last relation shows that the signs of $\frac{\partial^2 f}{\partial T^2}$ and $\frac{\partial^2 u}{\partial s^2}$ are opposite and more precisely that $\frac{\partial^2 f}{\partial T^2}\leq0$ due to the sign of $\frac{\partial^2 u}{\partial s^2}$ being positive, as demonstrated in Eq. \ref{eq:13}.

See that there is no mathematical reason to modify the second-order derivatives of $f$ in relation to the $v$ parameter, keeping it positive as it occurs to the energy function $u$, and so $\frac{\partial^{2} f}{\partial v^{2}}\geq0$. To compute this, the same way as previously presented demonstration to the $\frac{\partial^2 f}{\partial T^2}$, it is possible to compare the sign of $\frac{\partial^{2} f}{\partial v^{2}}$ with any known curvature of the energy function or any thermodynamic potentials eventually chosen. It is of complete generality the fact that Legendre transformation modify the concavity (or the curvature) of the original function in relation to the new parameters introduced on transformed function. It is relevant to stress that all other curvatures associated to the unmodified parameters by Legendre transformation keeps unaltered in the transformed function. For a mathematical general treatment about Legendre transformations see Refs. \cite{boas2006mathematical, callen1998thermodynamics}. 

The signs of the second-order derivatives for Helmholtz free energy are given by Eq. \ref{eq:20} and can be obtained using the same method and formalism already addressed above to the thermodynamic energy function. Mathematically the $f$ free energy is known as a saddle surface, as shown Fig. \ref{fig1}. If we employ the same method used for $u$ in the Helmholtz potential, in this case we need to set eigenvalues with opposite signs in canonical form of $f$ (similarly to the Eq. \ref{eq:6}). Remembering that we achieved $f$ by Legendre transformation on $s$ parameter of $u$ function whereby we introduce $T$ as variable of $f$. This surface of two variables $f$ has a maximum in relation to the temperature but a minimum in relation to the volume. 
\begin{flalign} 
\frac{\partial^{2} f}{\partial T^{2}}\leq 0\ \textrm{, } \frac{\partial^{2} f}{\partial v^{2}}\geq0 \textrm{ and }
\frac{\partial^{2} f}{\partial T^{2}}\frac{\partial^{2} f}{\partial v^{2}} - \Bigg(\frac{\partial^{2} f}{\partial T\partial v}\Bigg)^{2}\leq0
\label{eq:20}
\end{flalign}

In the same way as $f$, we can introduce the intensive parameter $p$ changing the extensive variable $v$ in $u$ in order to obtain other physical potential, i. e., the enthalpy $h$. This thermodynamic function is dependent on $s$ and $p$, $h=h(s,p)$. Remembering that $u=Tds-pdv$, then
\begin{flalign} 
d(pv)=pdv+vdp\nonumber \\
-pdv=-d(pv)+vdp.
\label{eq:21}
\end{flalign}

Substituting Eq. \ref{eq:21} into Eq. \ref{eq:16} and making some simple manipulations
\begin{flalign} 
du=Tds-d(pv)+vdp\nonumber \\
d(u+pv)=Tds+vdp\nonumber\\
dh=Tds+vdp,
\label{eq:22}
\end{flalign}
where 
\begin{flalign} 
h\equiv u +pv.
\label{eq:23}
\end{flalign}

As occured for Helmholtz potential, here the enthalpy potential is also mathematically a saddle surface due to the Legendre transformation being in an unique extensive parameter $v$. Further, $h$ keep unaltered with a  minimum in relation to the entropy $s$ but becomes a maximum on $p$, as direct consequence of the Legendre transformation that changes the variable $v$ by $p$. 

The signs of the second-order derivatives in enthalpy potential $h$ are given by Eq. \ref{eq:24}, obtained by analogy with the previous cases, Taylor's expansion, diagonalization of the appropriate quadratic form, etc.
\begin{flalign} 
\frac{\partial^{2} h}{\partial u^{2}}\geq0\ \textrm{, } \frac{\partial^{2} h}{\partial p^{2}}\leq0 \textrm{ and }
\frac{\partial^{2} h}{\partial u^{2}}\frac{\partial^{2} h}{\partial p^{2}} - \Bigg(\frac{\partial^{2} h}{\partial u\partial p}\Bigg)^{2}\leq0
\label{eq:24}
\end{flalign}

We previously determined that both thermodynamical potentials $f$ and $h$ are saddle surfaces. This occurs due to Legendre's transformation that inverses the curvature (or signs) in $f$ relative $T$ ($\frac{\partial^2 f}{\partial T^{2}}\leq0$) and $h$ relative $p$ ($\frac{\partial^{2}h}{\partial p^{2}}\leq0$). Imposing opposite signs in $\lambda_{1}$ and $\lambda_{2}$ in both canonical forms of the potentials $f$ and $h$ it is enough to obtain the set of Eqs. \ref{eq:20} and \ref{eq:24}.

If we combine the Eqs. \ref{eq:17} and \ref{eq:21} in Eq. \ref{eq:16} and after simple manipulations, it is possible to obtain another potential thermodynamic as a function of $T$ and $p$ namely the Gibbs potential $g=g(T,p)$. This is a mathematical process of double Legendre transformation that substitutes the extensive variables $s$ and $v$ in thermodynamic energy function by equivalent intensive variables $T$ and $p$.

The Legendre transformation completely change the curvature of the energy function, i. e, the second-order derivatives in relation to the $T$ and $p$ are negative now as well as the $g=g(T, p)$ becomes a surface of maximum. As it turns, the set of inequalities given Eq. \ref{eq:25} are representatives of the new signs of the second-order derivatives of the Gibbs potential. The double Legendre transformation change the curvature of the energy function relative its extensive parameters $s$ and $v$ (both positives in energy function) introducing new curvatures in relation to the intensive parameters $T$ and $p$ into Gibbs potential (both negatives in $g$ potential). We could simply have expanded the Gibbs potential in Taylor's series and obtained, as in previous cases, the same equation \ref{eq:25} through of the diagonalization of the quadratic form generated by expansion.
\begin{flalign} 
\frac{\partial^{2} g}{\partial T^{2}}\leq 0\ \textrm{, } \frac{\partial^{2} g}{\partial p^{2}}\leq 0 \textrm{ and }
\frac{\partial^{2} g}{\partial T^{2}}\frac{\partial^{2} g}{\partial p^{2}} - \Bigg(\frac{\partial^{2} g}{\partial T\partial p}\Bigg)^{2}\geq 0
\label{eq:25}
\end{flalign}

Furthermore, the Gibbs potential in this case is defined by
\begin{flalign} 
g\equiv u-Ts+pv.
\label{eq:26}
\end{flalign}

An interesting observation it is that thermodynamic functions are convex functions of their extensive variables (positive curvatures) and concave functions (negative curvatures) of their intensive variables. In entropy representation is possible to obtain potential functions given by expressions namely Massieu functions. It is easy to show that there is dependence between some Massieu functions and the potentials defined in the energy representation. But it is sufficient to present here  the energy representation of thermodynamic potentials to understand the stability of thermodynamic systems.

\section{A review on quantities of physical interest\label{review}} 

There are some physical quantities of interest, as coefficient of thermal expansion $\alpha$ at $p$ constant, specific heats at $v$ or $p$ constant ($c_v$ and $c_p$, respectively), isothermal compressibility $k_T$ ($T$ constant) and adiabatic compressibility $k_s$ ($s$ constant). These physical quantities are defined below \cite{callen1998thermodynamics,de2005termodinamica,wreszinski2003termodinamica}.
\begin{flalign} 
\alpha\equiv\frac{1}{v}\frac{\partial v}{\partial T}
\label{eq:27}
\end{flalign}

\begin{flalign} 
c_{v}\equiv T\frac{\partial s}{\partial T} \textrm{ , } c_{p}\equiv T\frac{\partial s}{\partial T}
\label{eq:28}
\end{flalign}

\begin{flalign} 
k_{T}\equiv-\frac{1}{v}\frac{\partial v}{\partial p} \textrm{ , } k_{s}\equiv-\frac{1}{v}\frac{\partial v}{\partial p}
\label{eq:29}
\end{flalign}

The thermal expansion $\alpha$ at $p$ constant is associated with changes in dimensions of physical systems due to temperature variations. This physical quantity is of great importance in physics because through its absolute value we can understand the behavior of materials on the macroscopic or microscopic scale when subjected to temperature changes. Besides, $\alpha$ can be positive or negative. For example, Yonn et. al. found negative thermal expansion coefficient of graphene measured by Raman spectroscopy graphene \cite{yoon2011negative}. For characterizing and understanding of the impact of temperature and internal stresses on the behavior of optical fibers, the physical quantity $\alpha$ is also useful \cite{cavillon2017additivity}.

The specific heats at constant pressure or volume are the quasi-static heat flux per mole that produces an increase of one unit at the temperature of a physical system. Through this physical quantity, we can understand the thermal properties of physical systems in several length scales (macroscale and microscale), as for example the measurements of heat capacity of the individual multiwalled carbon nanotubes where through of the thermal properties (heat capacity) it was possible to control and understand the performance and stability of nanotube devices \cite{kim2001thermal,dresselhaus1996science}. Specific heats are positive physical quantities related to the thermal stability of the system, as we show in next section. 

Isothermal and adiabatic compressibilities $k_{T}$ and $k_{s}$ are the fractional decrease in volume per unit of pressure. Through the obtention of these coefficients many works have been developed such as the importance of compressibility data for characterizing protein transitions \cite{taulier2002compressibility} as well as the study of compressibility of liquid metals \cite{marcus2017compressibility}. These physical quantities are positives and they are related to the mechanical stability of the system, as we show next.

There are relevant relationships between the above physical quantities presented \cite{callen1998thermodynamics, de2005termodinamica}. A formal comprehension of the physical origin of the quantities given by Eqs. \ref{eq:27}, \ref{eq:28} and \ref{eq:29} it is indispensable in any work of thermodynamics, especially for the appropriate interpretation of results in theoretical or experimental researches.

\section{The immediate consequences to the sign of physical quantities\label{consequences}} 

Let us start by Helmholtz potential and investigating the possible physical consequences on stability associated with the curvature of the mentioned potential. An infinitesimal differential of $f$ can be obtained as follows
\begin{flalign}
f=u-Ts=f(T,v)\nonumber\\
df=\frac{\partial f}{\partial T}dT+\frac{\partial f}{\partial v}dv.
\label{eq:30}
\end{flalign}

By comparing Eq. \ref{eq:30} with Eq. \ref{eq:18} it is possible to obtain
\begin{flalign}
-s=\frac{\partial f}{\partial T},
\label{eq:31}
\end{flalign}
and
\begin{flalign}
-p=\frac{\partial f}{\partial v}.
\label{eq:32}
\end{flalign}

If we take the derivation side by side of Eq. \ref{eq:31} in relation to the $T$ considering $v$ constant 
\begin{flalign}
-\frac{\partial s}{\partial T} =\frac{\partial^2 f}{\partial T^2}.
\label{eq:33}
\end{flalign}
See that the left side of the expression above is relationed to the specific heat at $v$ constant and the sign of the second-order derivatives of Eq. \ref{eq:33} can be checked by comparing with Eq. \ref{eq:20}, and so
\begin{flalign}
-\frac{\partial s}{\partial T} =\frac{\partial^2 f}{\partial T^2}\leq0.
\label{eq:34}
\end{flalign}
From definition of the specific heat at $v$ constant given by \ref{eq:28}, we can obtain
\begin{flalign}
-\frac{c_{v}}{T} \leq0 \Longrightarrow  \frac{c_{v}}{T} \geq0 \Longrightarrow \boxed{c_{v}\geq0}
\label{eq:35}
\end{flalign}

As the absolute temperature is always positive it is possible to conclude that the physical quantity $c_{v}$ (specific heat at constant volume) is also necessarily positive,  $c_{v}\geq0$. The specific heat must be imagined as the necessary quantity of heat to increase or decrease the temperature of a physical system (see previous section). Furthermore, a positive increment of temperature $dT>0$ in system is associated with the transfer of heat into the system, also positive. On the other hand, a decrease in temperature $dT<0$ is associated with the outing heat of a system, also negative. In both cases the reason between heat differences and temperature is positive and intimatily relationed to the stability of the physical system considered. Conversely, a negative specific heat would imply in an inexistent physical situation because we would have a system capable of receiving some quantity of heat (postive) and decreasing its temperature, and so $dT<0$. If this non-physical situation could happen, the system would absorb indefinitely heat due to the difference of temperature between environment and system, existent because the system always decreases its temperature when receiving the environment heat. There is another non-physical situation with negative specific heat in the hypothetical situation in which the system loses heat but increases its temperature.

See that the sign of $\frac{\partial^2{f}}{\partial T^2}\leq0$ has been directly relationed to the sign of the specific heat at $v$ constant. This appears as an immidiate consequence of the curvature in Helmholtz potential. What is more, the physical stability naturally emerges of the mathematical features of $f$. For completeness, the princicle of minimum energy provides the sign of $c_{v}$ of a fundamental point of view (remembering that $f$ is obtained of $u$). We can also observe that the mathematical procedure of Legendre transformation only changes the sign of curvature of the Helmoltz potential in relation to the extensive parameter $s$ present in energy representation. 

We could find equal results to the positive sign of the specific heat using the signs of second-order derivatives of the thermodynamic energy function instead of using Hel\-mholtz potential. See that in Eq. \ref{eq:13}
\begin{flalign}
\frac{\partial^{2}u }{\partial s^{2}} \geq0.
\label{eq:36}
\end{flalign}
Then, we know that the temperature definition is given by
\begin{flalign}
T=\frac{\partial u }{\partial s}. 
\label{eq:37}
\end{flalign}

If we take the derivation of Eq. \ref{eq:37} side by side in relation to the $s$ entropy
\begin{flalign}
\frac{\partial T}{\partial s}=\frac{\partial^{2} u }{\partial s^{2}}\geq0,
\label{eq:38}
\end{flalign}
And combining Eq. \ref{eq:38} and the definition of specific heat given by Eq. \ref{eq:28}, we have
\begin{flalign}
\frac{T}{c_{v}}=\frac{\partial^{2} u }{\partial s^{2}}\geq0 \Longrightarrow c_{v}\geq0.
\label{eq:39}
\end{flalign}
The expression above leads to a positive specific heat ($c_{v}\geq0$) because the absolute temperature is positive, as already obtained in Eq. \ref{eq:35}. Notice that we have to the same conclusions using different representations. It is necessary to just choose the appropriate potential; and by analyzing of the second-order derivatives, we obtain relevant relationships between the physical quantities presented in previous section and the signs of the quantities that have direct implication on physical stability (see more details on relations between the physical quantities in Ref. \cite{callen1998thermodynamics}). Here, we are interested in showing that the signs of the physical quantities can be found in the curvatures of the potentials, etc.

Let us analyze now the curvature of the Gibbs potential, in particular, the first expression in Eq. \ref{eq:25}. But before that, it is interesting to remind that  Eq. \ref{eq:26} is obtained by double Legendre transformation in extensive variables $s$ and $v$ of $u=u(s,v)$, introducing $T$ and $p$ in $g$ function. A differential element $dg$ of Gibbs potential is given by
\begin{flalign}
dg=-sdT+vdp.
\label{eq:40}
\end{flalign}
As the $g=g(T,p)$, we have on the other hand
\begin{flalign}
dg=\frac{\partial g}{\partial T}dT+\frac{\partial g}{\partial p}dp.
\label{eq:41}
\end{flalign}
By comparing Eqs. \ref{eq:40} and \ref{eq:41}, it is possible to concludes that,
\begin{flalign}
-s=\frac{\partial g}{\partial T},
\label{eq:42}
\end{flalign}
and 
\begin{flalign}
v=\frac{\partial g}{\partial p}.
\label{eq:43}
\end{flalign}

Taking the derivation of Eq. \ref{eq:42} side by side in relation to the temperature at $p$ constant
\begin{flalign}
-\frac{\partial s}{\partial T}=\frac{\partial^2 g}{\partial T^2}.
\label{eq:44}
\end{flalign}
From definition of specific heat at $p$ constant given by Eq. \ref{eq:28} and looking at the sign of the first expression in Eq. \ref{eq:25}, we have
\begin{flalign}
-\frac{c_{p}}{T}=\frac{\partial^2 g}{\partial T^2}\leq0\nonumber\\
\frac{c_{p}}{T}\geq0 \Longrightarrow \boxed{c_{p}\geq0}.
\label{eq:45}
\end{flalign}

Notice that as well as for the specific heat at $v$ constant, we have here that specific heat at $p$ constant is also positive, $c_{p}\geq0$. This is directly associated with the stability of the physical system, as discussed above on $c_{v}$. It is relevant to emphasize that the choice of the appropriate thermodynamic potential is definitely associated with the success in finding some stability condition. As the specific heat at $p$ constant involves derivatives of $s$ in relation to $T$, there is an efficient rule that we can consider, that is,  immediately look for Gibbs potential due to $g=g(T,p)$ being a funcion on $T$ and $p$ (that appear when considering specific heat definition). But this is not the only reason to find some stability condition, as we show after to isothermal compressibility $k_{T}$. In this case, we naturally look for the second-order derivatives of $g$ involving $T$ and $p$ parameters, i. e., $\frac{\partial^2 g}{\partial p^2}\leq0$, and this is possible by using the same procedure already discussed.

As previously stated, the appropriate choice of some thermodynamic potential is enough, and not necessarily the potential that involves $T$ and $p$ parameters. Notice that, by choice the expression given by Eq. \ref{eq:32} and, therefore, Helmholtz's potential, we can compute the sign of $k_{T}$ as shown below.
\begin{flalign}
-p=\frac{\partial f}{\partial v}
\label{eq:46}
\end{flalign}
By derivation of the left and right sides in relation to the volume keeping temperature constant
\begin{flalign}
\frac{\partial p}{\partial v}=-\frac{\partial^2 f}{\partial v^2}.
\label{eq:47}
\end{flalign}
From Eq. \ref{eq:29} we have
\begin{flalign}
\frac{\partial p}{\partial v}=-\frac{1}{v k_{T}},
\label{eq:48}
\end{flalign}
And combining Eqs. \ref{eq:47} and \ref{eq:48} as well as by recalling the expression \ref{eq:20} that provides a positive sign of the second-order derivatives in relation to the extensive parameter, we conclude
\begin{flalign}
-\frac{1}{v k_{T}}=-\frac{\partial^2 f}{\partial v^2}\nonumber\\
\frac{1}{v k_{T}}=\frac{\partial^2 f}{\partial v^2}\geq0 \Longrightarrow \boxed{k_{T}\geq0}.
\label{eq:49}
\end{flalign}
As in previous cases the curvature or signal of the appropriate potential leads to a relevant relation for the signal of physical quantity of interest, $k_{T}\geq0$ in this case. It is still intuitive to notice in Eq. \ref{eq:29} that increments of pressure in the system leads to decrease in volume due to the ever positive isothermal compressibility. Here it is interesting to observe that from a fundamental point of view, the sign of $k_{T}$ emerges from the principle of minimum energy or its entropic equivalent. Remembering that Helmholtz is only an alternative representation that provides the physics of the system in pratical experiments (see the importance of representations in Ref. \cite{callen1998thermodynamics}). In more elementary courses the negative sign that appears in Eq. \ref{eq:29} is adjusted without discussing the formal aspects of the potential's curvature, clearly demonstrated here.

Let us investigate if there is similar stability condition for the adiabatic compressibility $k_{s}$, as calculated for $k_{T}$. We have in this physical quantity a derivation in relation to $p$ at $s$ constant. Following the same protocol used to investigating the signs of the physical quantities already addressed; it is convinient to consider now Helmholtz's potential $h$ since it is a function on $s$ and $p$. Although this rule is not the only method. The same way, we could use other potentials to obtain some stability condition for the $k_{s}$.

The expression given by Eq. \ref{eq:23} (enthalpy potential) is computed to be a function on $s$ and $p$ applying an viable Legendre transformation in $v$, and introducing $p$ in the energy formalism. See that,
\begin{flalign}
h=u+pv=h(s,p)\nonumber\\
dh=\frac{\partial h}{\partial s}ds+\frac{\partial h}{\partial p}dp.
\label{eq:50}
\end{flalign}

By comparing Eq. \ref{eq:50} with Eq. \ref{eq:22} it is possible to conclude
\begin{flalign}
T=\frac{\partial h}{\partial s},
\label{eq:51}
\end{flalign}
and 
\begin{flalign}
v=\frac{\partial h}{\partial p}.
\label{eq:52}
\end{flalign}
Taking the derivation of both sides in  Eq. \ref{eq:52} above in relation to $p$ at $s$ constant
\begin{flalign}
\frac{\partial v}{\partial p}=\frac{\partial^2 h}{\partial p^2},
\label{eq:53}
\end{flalign}
and rewriting conveniently the definition of $k_{s}$ in Eq. \ref{eq:29}, we obtain
\begin{flalign}
-vk_{s}=\frac{\partial^2 h}{\partial p^2} \Longrightarrow \boxed{k_{s}\geq0}.
\label{eq:54}
\end{flalign}

As well as $k_{T}$ the positive physical quantity $k_{s}$ is associated with the mechanical stability of a physical system. Besides, it is directly related to the curvature of the appropriate potential, $h$ enthalpy this time. The positive  compressibility coefficients are related to the mechanical stability of the system and emerges as an immediate consequence of curvature of the potentials described above. 

The thermal expansion coefficient $\alpha$ does not to have a positive defined sign that can be obtained from some appropriate thermodynamic potential. In the well-known case of the water, for example, the volume increases when temperature decreases below at $4^{o}C$, and $\alpha$ is negative in this regime.

There are some known relationships between those physical quantities, as commonly demonstrated in thermodynamics books (see Refs. \cite{callen1998thermodynamics, de2005termodinamica,wreszinski2003termodinamica}) and written below
\begin{flalign}
c_{p}=c_{v}+\frac{Tv\alpha^{2}}{k_T}
\label{eq:55}
\end{flalign}
\begin{flalign}
\frac{c_{p}}{c_{v}}=\frac{k_T}{k_s}.
\label{eq:56}
\end{flalign}
The set of expressions given by the Eqs. \ref{eq:55} and \ref{eq:56} shown that $c_{p}\geq c_{v}$ and $k_{T}\geq k_{s}$. It is not the purpose of this work demonstrates the equations \ref{eq:55} and \ref{eq:56} that can be obtained by reduction of thermodynamic derivatives and by using Maxwell's relations appropriately chosen (see more details in Ref. \cite{callen1998thermodynamics}).

\section{Conclusions\label{summary}}

It is worthy of emphasis that the present paper demonstrates the power of superposition of mathematical and physical aspects in order to understand the stability of thermodynamic systems. We have clearly proven that the thermal and mechanical stability of physical systems are directly associated with the curvature of the considered thermodynamic potentials, including energy representation. As the conditions of stability were directly obtained from the curvature of the potentials, we conclude that of a fundamental point from view the stability conditions result from the minimum energy principle. The thermodynamic potentials emerges as a useful representation to address more relalistic or experimental problems, and its second-order derivatives are obtained by diagonalization of the expanded quadratic forms through Taylor's series. In particular, we consistently demonstrated formal aspects of the thermodynamic theory with important consequences on thermal and mechanical stability. The used mathematics mode of quadratic  diagonalization forms to obtain the thermodynamic potentials is also relevant as a support for postulational thermodynamics approach of traditional undergraduate and postgraduate courses, particularly courses that require fundamental knowledments on the stability.

\section{Acknowledgments}

The authors gratefully acknowledge the support provided by Brazilian Agencies CAPES, CNPQ and FAPESP. We would like to thank the following for their kind support: Piau\'i Federal Institute, S\~ao Raimundo Nonato campus; Professor Jos\'e Pimentel de Lima (Departament of Physics -- UFPI), for the fundamental basis in Physics and the great lectures on Thermodynamics during the years as an undergraduate that we shall keep for the rest of ours lives. In one way, his name can be seen on all the pages of this paper. Finally, we would like to show our deepest appreciation for our friend and colleague Israel A. C. Noletto, for proofreading this paper.

\bibliographystyle{unsrt}  


\bibliographystyle{elsarticle-num}
\bibliography{references}

\end{document}